\documentclass[prb,cha,twocolumn,superscriptaddress,preprintnumbers,showpacs, amsmath,amssymb]{revtex4-1}
\usepackage{bm}
\usepackage[colorlinks=true,linkcolor=blue,citecolor=blue]{hyperref}
\usepackage{amsmath}
\usepackage{amssymb}
\usepackage{amsthm}
\usepackage{amsfonts}
\usepackage{enumerate}
\usepackage{latexsym}
\usepackage{ifpdf}
\usepackage{graphicx}
\usepackage{makeidx}
\expandafter\ifx\csname package@font\endcsname\relax\else
 \expandafter\expandafter
 \expandafter\usepackage
 \expandafter\expandafter
 \expandafter{\csname package@font\endcsname}
 \fi
\usepackage{color}
\usepackage{times}

\begin{document}

\title{Metallic interfaces in a CaTiO$_3$/LaTiO$_3$ heterostructure}

\author{Shaozhu Xiao}
\affiliation{Ningbo Institute of Materials Technology and Engineering, Chinese Academy of Sciences, Ningbo 315201, China}
\author{Fangdi Wen}
\affiliation{Department of Physics and Astronomy, Rutgers University, Piscataway, New Jersey 08854, USA}
\author{Xiaoran Liu}
\affiliation{Department of Physics and Astronomy, Rutgers University, Piscataway, New Jersey 08854, USA}
\author{M. Kareev}
\affiliation{Department of Physics and Astronomy, Rutgers University, Piscataway, New Jersey 08854, USA}
\author{Yang Song}
\affiliation{Ningbo Institute of Materials Technology and Engineering, Chinese Academy of Sciences, Ningbo 315201, China}
\author{Ruyi Zhang}
\affiliation{Ningbo Institute of Materials Technology and Engineering, Chinese Academy of Sciences, Ningbo 315201, China}
\author{Yujuan Pei}
\affiliation{Ningbo Institute of Materials Technology and Engineering, Chinese Academy of Sciences, Ningbo 315201, China}
\author{Jiachang Bi}
\affiliation{Ningbo Institute of Materials Technology and Engineering, Chinese Academy of Sciences, Ningbo 315201, China}
\author{Shaolong He}
\affiliation{Ningbo Institute of Materials Technology and Engineering, Chinese Academy of Sciences, Ningbo 315201, China}
\author{Yanwei Cao}
\email{ywcao@nimte.ac.cn}
\affiliation{Ningbo Institute of Materials Technology and Engineering, Chinese Academy of Sciences, Ningbo 315201, China}
\author{J. Chakhalian}
\affiliation{Department of Physics and Astronomy, Rutgers University, Piscataway, New Jersey 08854, USA}

\date{\today}

\begin{abstract}

Almost all oxide two-dimensional electron gases are formed in SrTiO$_3$-based heterostructures and the study of non-SrTiO$_3$ systems is extremely rare. Here, we report the realization of a two-dimensional electron gas in a CaTiO$_3$-based heterostructure, CaTiO$_3$/LaTiO$_3$, grown epitaxially layer-by-layer on a NdGaO$_3$ (110) substrate via pulsed laser deposition. The high quality of the crystal and electronic structures are characterized by in-situ reflection high-energy electron diffraction, X-ray diffraction, and X-ray photoemission spectroscopy. Measurement of electrical transport validates the formation of a two-dimensional electron gas in the CaTiO$_3$/LaTiO$_3$ superlattice. It is revealed the room-temperature carrier mobility in CaTiO$_3$/LaTiO$_3$ is nearly 3 times higher than in CaTiO$_3$/YTiO$_3$, demonstrating the effect of TiO$_6$ octahedral tilts and rotations on carrier mobility of two-dimensional electron gases. Due to doped CaTiO$_3$ being an A-site polar metal, our results provide a new route to design novel A-site two-dimensional polar metals. 

\end{abstract}

\maketitle

\newpage

The entanglement among charge, spin, orbital, and structural degrees of freedom in transition metal oxides leads to a great number of emergent phenomena, one remarkable example of which is the formation of a two-dimensional electron gas (2DEG) at the interface of two insulating compounds, e.g., SrTiO$_3$/LaAlO$_3$,\cite{Nature-2004-Hwang,APL-2002-Tokura} SrTiO$_3$/$\gamma$-Al$_2$O$_3$, \cite{NC-2013-Chen,NPJQM-2016-Cao} SrTiO$_3$/$R$TiO$_3$ ($R$ = La, Gd, Nd), \cite{Nature-2002-Hwang,JJAP-2004-KS,APL-2011-Stemmer,APL-2014-BJ} SrTiO$_3$/LaVO$_3$, \cite{PRL-2007-Hwang} SrTiO$_3$/LaGaO$_3$,\cite{APL-2010-PP} and SrTiO$_3$/DyScO$_3$. \cite{APL-2011-Li} Quantum many-body effects at the interfaces of transition metal oxides benefit the formation of two-dimensional polar metals, magnetic 2DEGs, and the coexistence of superconductivity and ferromagnetism.\cite{NC-2018-Cao,PRL-2016-Cao,APL-2018-Wen,PRL-2011-DD,NP-2011-JB,NP-2011-LL} On the other hand, strong correlations result in the room-temperature carrier mobilities of SrTiO$_3$-based 2DEGs being ultra-low (0.1--10~cm$^2$V$^{-1}$s$^{-1}$), \cite{ARMR-2014-Stemmer} which hinders their practical applications for advanced electronic devices. As seen, almost all perovskite oxide 2DEGs are found in SrTiO$_3$-based heterostructures, the two unique exceptions being KTaO$_3$/LaTiO$_3$ and CaTiO$_3$/YTiO$_3$. \cite{APLM-2015-Ahn,APL-2015-Liu} Therefore, the study of non-SrTiO$_3$ systems is thus essential, but essentially lacking thus far.

\begin{figure}[b]
	\includegraphics[width=\columnwidth]{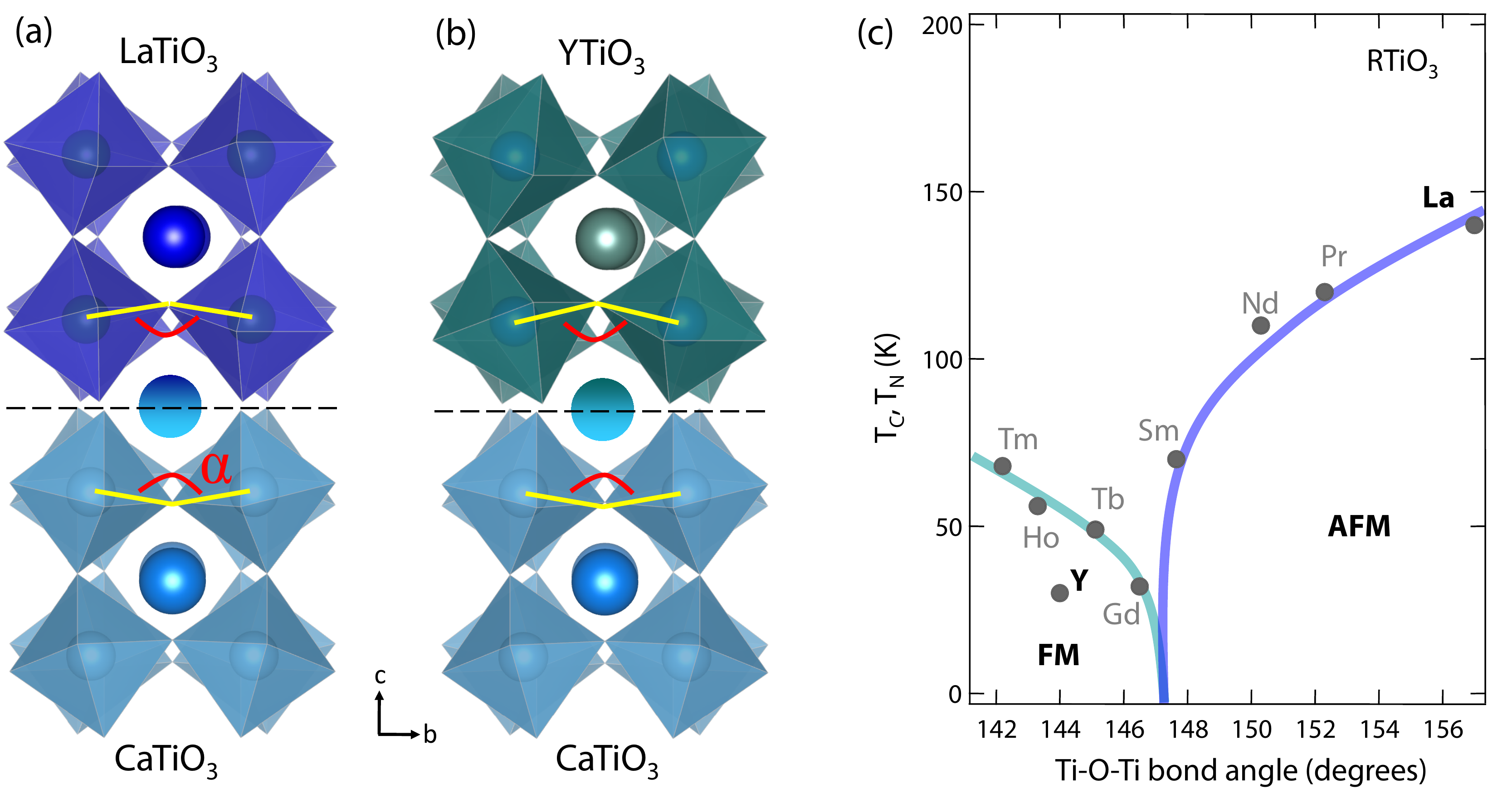}
	\caption{\label{fig1}Schematic of (a) CTO/LTO and (b) CTO/YTO heterostructures. The yellow lines mark the Ti--O--Ti bond angle $\alpha$. (c) Adapted phase diagram of $R$TiO$_3$ ($R$ a rare-earth element) from Ref.~\onlinecite{PRB-1997-Tokura,JSSC-1979-Greedan}. With decreasing Ti--O--Ti bond angle from La to Tm, the magnetic ground state changes from antiferromagnetic (AFM) to ferromagnetic (FM).}
\end{figure}

To address this issue, we focus the study of CaTiO$_3$-based heterostructures. Here, we take the heterostructure CaTiO$_3$/LaTiO$_3$ [CTO/LTO, see Fig.~\ref{fig1}(a)] as a prototype to investigate the effect of TiO$_6$ octahedra on carrier mobility by comparing its properties with the heterostructure CaTiO$_3$/YTiO$_3$ [CTO/YTO, see Fig.~\ref{fig1}(b)].\cite{APL-2015-Liu} As seen in Fig.~\ref{fig1}(c), in bulk LTO the Ti--O--Ti bond angle $\alpha$ is 157$^\circ$, whereas it is only 144$^\circ$ in YTO, which results in a larger conduction band width of LTO (2.45~eV) than of YTO (2.04~eV).\cite{PRB-1997-Tokura} The Ti--O--Ti bond angle can play a key role in determining the electrical transport properties,\cite{ACSAMI-2015-TS} and is expected to lead to a higher carrier mobility in CTO/LTO than in CTO/YTO. Here, the material LTO is carefully selected for the reason that its bond angle is very close to that of bulk CTO ($\sim$ 156$^\circ$).\cite{JSSC-1979-Greedan,JSSC-2003-Eng} Moreover, the lattice parameters of bulk CTO at room temperature $a$ = 5.38~\AA, $b$ = 5.44~\AA, $c$ = 7.64~\AA\ (in the orthorhombic setting) are well-matched with the lattice parameters of the NdGaO$_3$ (NGO) substrate $a$ = 5.43~\AA, $b$ = 5.50~\AA, $c$ = 7.71~\AA.\cite{JSSC-2005-Ali}

In this Letter, we report the layer-by-layer synthesis of high quality CTO/LTO superlattices by pulsed laser deposition (PLD). The high quality of the crystal and electronic structures are characterized by reflection high-energy electron diffraction (RHEED), X-ray diffraction (XRD), and X-ray photoemission spectroscopy (XPS). The formation of a 2DEG in CTO/LTO is verified with temperature-dependent electrical transport, revealing a room temperature carrier mobility in CTO/LTO nearly triple that in more-strongly-distorted CTO/YTO. 

\begin{figure}[]
	\includegraphics[width=\columnwidth]{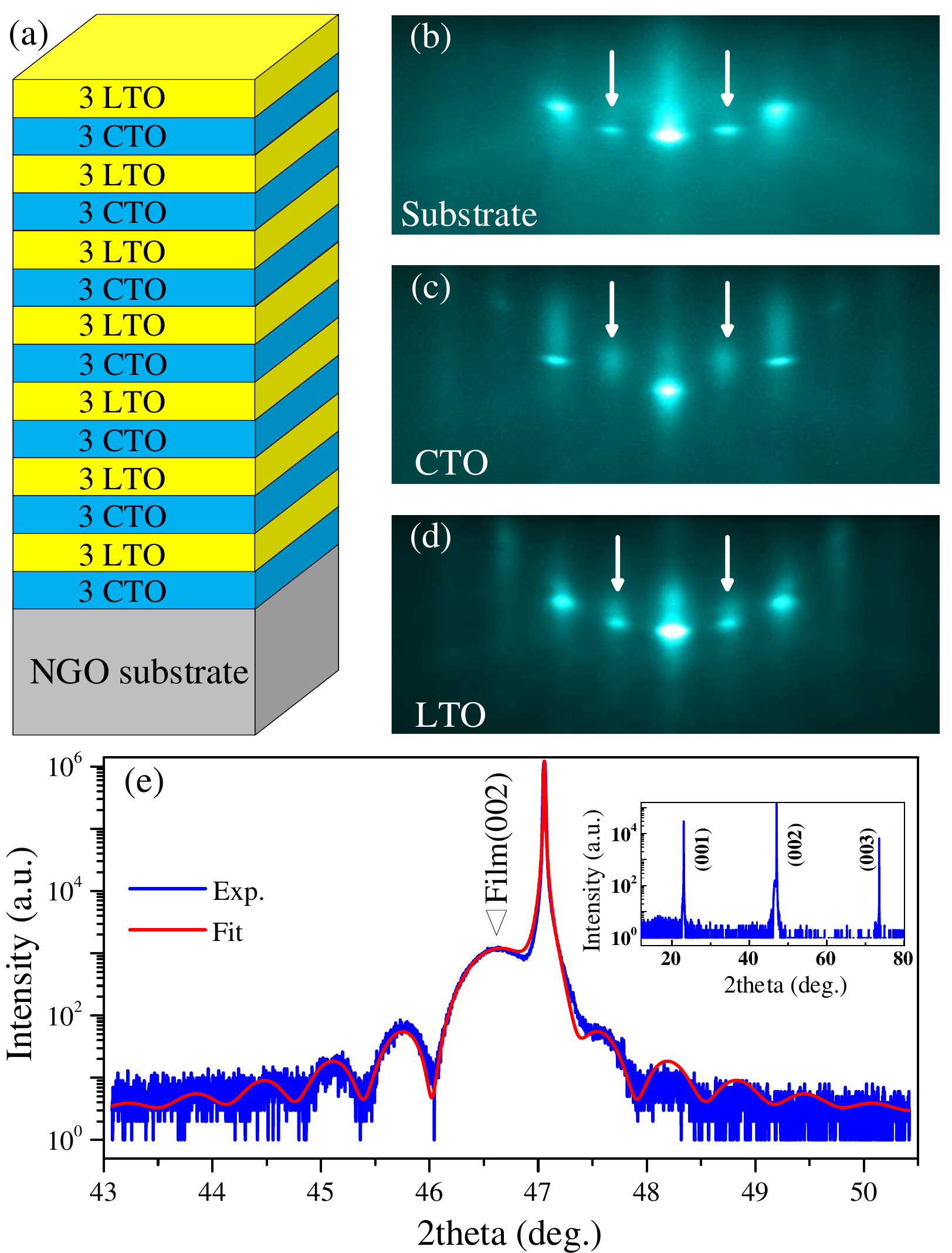}
	\caption{\label{fig2}(a) Schematic of 3CTO/3LTO superlattice on an NGO(110) substrate. (b-d) RHEED patterns of NGO substrate, CTO layer, and LTO layer during growth, respectively. The white arrows identify fractional reflections in the cubic setting, indicating orthorhombic symmetry. (e) X-ray diffraction $\omega$--$2\theta$ scan curve of a 3CTO/3LTO superlattice on an NGO substrate around the (002) reflection. The triangle indicates the (002) peak of the film. Inset: wide range $\omega$--$2\theta$ scan curve.}
\end{figure}

\begin{figure}[]
	\includegraphics[width=\columnwidth]{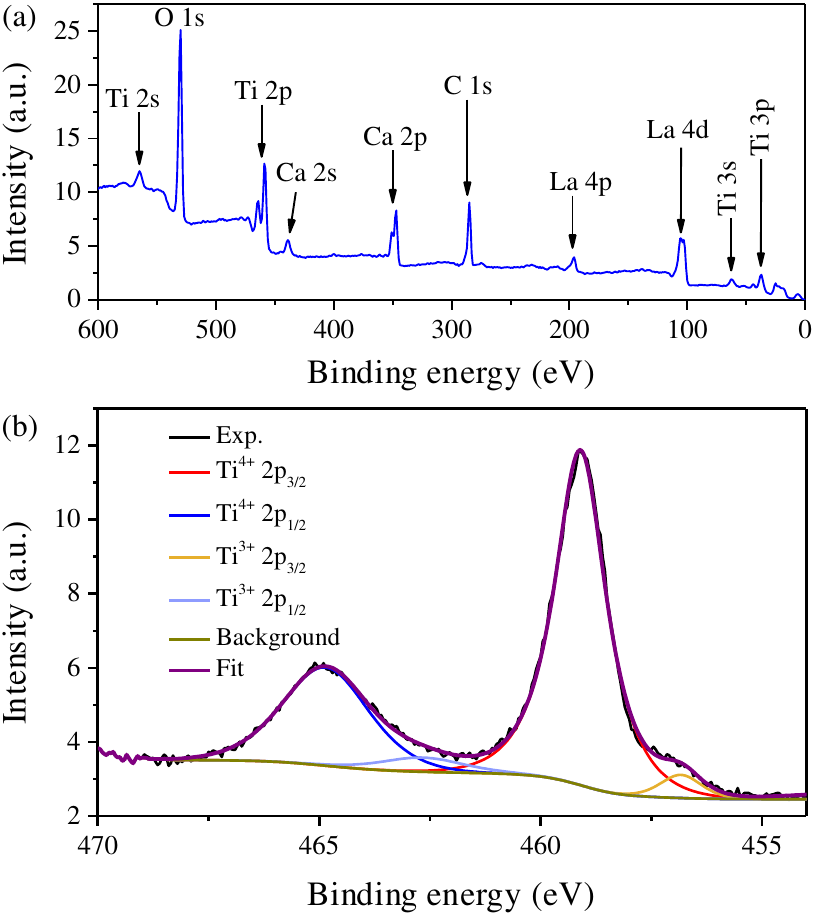}
	\caption{\label{fig3}XPS spectra of 3CTO/3LTO superlattice. The spectra was obtained along the normal direction of sample. (a) Spectra from 0 to 600~eV binding energy. (b) Ti 2$p$ XPS at room temperature.}
\end{figure}

The superlattices [3~u.c.~CTO/3~u.c.~LTO]$_7$ [u.c. = pseudocubic unit cells, see Fig.~\ref{fig2}(a)] were synthesized on (110)-oriented NGO (orthorhombic notation) single crystal substrates (5 $\times$ 5 $\times$ 0.5 mm$^3$) by PLD using a KrF excimer laser operating at $\lambda$ = 248~nm and a pulse rate of 2~Hz. The details of growth parameters can be found in our previous reports on CTO films and CTO/YTO heterostructures.\cite{APL-2015-Liu,APL-2016-Cao} The high quality of the films was confirmed by both \textit{in-situ} RHEED and \textit{ex-situ} XRD (Cu K$_{\alpha1}$, $\lambda$ = 1.5406~\AA). To investigate the electronic structure, composition, and valence states of the films, XPS (monochromated Al K$_{\alpha}$ radiation, h$\nu$ = 1486.6~eV) was performed. Sheet resistances and Hall effect were measured between 2 and 300~K using a Physical Property Measurement System (Quantum Design PPMS EverCool-$\rm\uppercase\expandafter{\romannumeral2}$) in the van der Pauw geometry. 

Figure~\ref{fig2} shows the crystal structure of the 3CTO/3LTO film. As seen in Fig.~\ref{fig2} (b-d), sharp RHEED patterns from the CTO and LTO layers during epitaxy indicate layer-by-layer growth of 3CTO/3LTO superlattices. The observation of clear half order peaks [white arrows in Fig.~\ref{fig2} (b-d)] confirms the orthorhombic symmetry of the film. To further investigate the structural details of the sample, we performed XRD $\omega$--$2\theta$ scan measurements on the sample. As seen from the wide range $\omega$--$2\theta$ scan curve shown in the inset of Fig.~\ref{fig2} (e), no secondary phase is observed. Around the sharp peak from the NGO substrate at $2\theta=47.06^\circ$, a broad film peak [indicated by a triangle in Fig.~\ref{fig2} (e)] with Kiessig fringes can be observed, further confirming the crystallinity of the superlattices. By fitting of the Kiessig fringes [see Fig.~\ref{fig2} (e)], the thickness of the film is estimated to be 15.2~nm, consistent with the expected value 16.2~nm.

Next, to investigate the chemical composition and valence states of the film, we carried out XPS at acceptance angles of 90$^\circ$ with the greatest penetration depth and 45$^\circ$ with dominant surface contribution. Figure~\ref{fig3} (a) shows a wide-energy XPS spectrum from 0 to 600 eV. As seen, no discernable impurity signal is observed besides adsorbed carbon on the film surface. Next, we focus on the charge states of titantum in 3CTO/3LTO. As displayed in Fig.~\ref{fig3} (b), the dominant feature of the Ti 2$p$ XPS spectra is doublet states. The two main peaks at 464.9~eV and 459.1~eV are assigned to Ti$^{4+}$ 2p$_{1/2}$ and Ti$^{4+}$ 2p$_{3/2}$ peaks, respectively. Shoulders at the low-binding-energy side of each peak represent the contribution of Ti$^{3+}$ with the Ti$^{3+}$ 2p$_{1/2}$ peak at about 462.6~eV and Ti$^{3+}$ 2p$_{3/2}$ at 456.8~eV. \cite{APL-2018-Wen,ASS-2018-Mu} Comparing with the data at acceptance angle 90$^\circ$, the spectral weight of Ti$^{3+}$ is decreased at acceptance angle 45$^\circ$ (more surface sensitive), indicating interfacial charge transfer from LTO to CTO sides. 

\begin{figure}[htb]
	\includegraphics[width=\columnwidth]{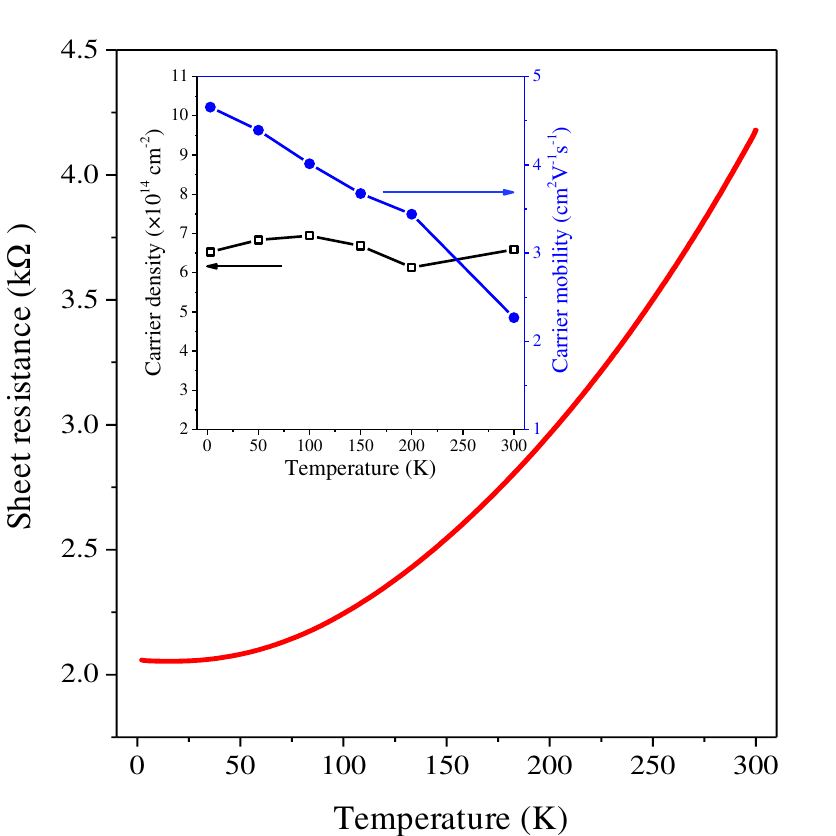}
	\caption{\label{fig4} Temperature-dependent sheet resistance of 3CTO/3LTO. Inset: carrier density and carrier mobility.}
\end{figure}

To verify the formation of a 2DEG at the CTO/LTO interfaces, temperature-dependent resistance measurements were carried out. As shown in Fig.~\ref{fig4}, on cooling from 300 to 3~K, the sheet resistance per interface falls from 4.2 to 2.1~k$\Omega$ per square. Considering that CTO, LTO, and the NGO substrate are all insulating, and that the NGO/CTO interface is also insulating,\cite{APL-2015-Liu} the CTO/LTO interfaces are the only possible conducting channels, confirming the formation of 2DEGs at the CTO/LTO interfaces. Next, to establish the type of charge carriers and estimate the carrier density, we measured the Hall resistance in vertical external magnetic fields from $-3$ to 3~T. The Hall resistance is linear in applied magnetic field, and the slope indicates the carriers in CTO/LTO are electron-like. As seen in the inset of Fig.~\ref{fig4}, the carrier density is almost temperature-independent at $6.5\times10^{14}$~cm$^{-2}$, whereas the carrier mobility increases from 2.3 to 4.7~cm$^2$V$^{-1}$s$^{-1}$ from 300 to 3~K, values $\sim$3 times larger than those of the previously-reported CTO/YTO superlattice. \cite{APL-2015-Liu} Based on the tight-binding calculations with the effect of TiO$_6$ distortion,\cite{PRB-1997-Tokura} the conduction band width of LTO (2.45~eV) is 19.6\% larger than that of YTO (2.04~eV), a consequence of the larger Ti-O-Ti bond angle in LTO than in YTO. Therefore, the higher carrier mobility in CTO/LTO than CTO/YTO most likely results from the larger band width of LTO.

In summary, we synthesized high quality CTO/LTO superlattices by PLD and characterized their crystal and electronic structures with RHEED, XRD, and XPS. Temperature-dependent electrical transport verifies the formation of a 2DEG. The carrier mobility in CTO/LTO is nearly 3 times higher than that in more-strongly-distorted CTO/YTO, resulting from weaker octahedral distortion of TiO$_6$ in LTO probably. Our results provide a route to design novel A-site two-dimensional polar metals.

This work is supported by the National Natural Science Foundation of China (Grant No. 11874058). Y. C. is supported by the Pioneer Hundred Talents Program of the Chinese Academy of Sciences and the Ningbo 3315 Innovation Team. J. C. acknowledges support from the Gordon and Betty Moore Foundation EPiQS Initiative through Grant No. GBMF4534. F.W. was supported by the Claud Lovelace Graduate Fellowship.

\newpage


\begin{thebibliography}{999}

\bibitem{Nature-2004-Hwang}
A. Ohtomo and H. Y. Hwang, Nature \textbf{427}, 423 (2004).

\bibitem{APL-2002-Tokura}
H. Yamada, M. Kawasaki, Y. Ogawa, and Y. Tokura, Appl. Phys. Lett. \textbf{81}, 4793 (2002).

\bibitem{NC-2013-Chen}
Y. Z. Chen, N. Bovet, F. Trier, D. V. Christensen, F. M. Qu, N. H. Andersen, T. Kasama, W. Zhang, R. Giraud, J. Dufouleur, T. S. Jespersen, J. R. Sun, A. Smith, J. Nygård, L. Lu, B. Büchner, B. G. Shen, S. Linderoth, and N. Pryds, Nat. Commun. \textbf{4}, 1371 (2013).

\bibitem{NPJQM-2016-Cao}
Y. Cao, X. Liu, P. Shafer, S. Middey, D. Meyers, M. Kareev, Z. Zhong, J. Kim, P. Ryan, E. Arenholz, and J. Chakhalian, npj Quantum Mater. \textbf{1}, 16009 (2016).

\bibitem{Nature-2002-Hwang}
A. Ohtomo, D. A. Muller, J. L. Grazul, and H. Y. Hwang, Nature \textbf{419}, 378 (2002).

\bibitem{JJAP-2004-KS}
K. Shibuya, T. Ohnishi, M. Kawasaki, H. Koinuma, and M. Lippmaa, Jpn. J. Appl. Phys. \textbf{43}, L1178 (2004).

\bibitem{APL-2011-Stemmer}
P. Moetakef, T. A. Cain, D. G. Ouellette, J. Y. Zhang, D. O. Klenov, A. Janotti, C. G. Van de Walle, S. Rajan, S. J. Allen, and S. Stemmer, Appl. Phys. Lett. \textbf{99}, 232116 (2011).

\bibitem{APL-2014-BJ}
P. Xu, D. Phelan, J. S. Jeong, K. A. Mkhoyan, and B. Jalan, Appl. Phys. Lett. \textbf{104}, 082109 (2014).

\bibitem{PRL-2007-Hwang}
Y. Hotta, T. Susaki, and H. Y. Hwang, Phys. Rev. Lett. \textbf{99}, 236805 (2007).

\bibitem{APL-2010-PP}
P. Perna, D. Maccariello, M. Radovic, U. Scotti di Uccio, I. Pallecchi, M. Codda, D. Marr, C. Cantoni, J. Gazquez, M. Varela, S. J. Pennycook, and F. M. Granozio, Appl. Phys. Lett. \textbf{97}, 152111 (2010).

\bibitem{APL-2011-Li}
D. F. Li, Y. Wang, and J. Y. Dai, Appl. Phys. Lett. \textbf{98}, 122108 (2011).

\bibitem{NC-2018-Cao}
Y. Cao, Z. Wang, S. Park, Y. Yuan, X. Liu, S. Nikitin, H. Akamatsu, M. Kareev, S. Middey, D. Meyers, P. Thompson, P. J. Ryan, P. Shafer, A. N'Diaye, E. Arenholz, V. Gopalan, Y. Zhu, K. M. Rabe, and J. Chakhalian, Nat. Commun. \textbf{9}, 1547 (2018).

\bibitem{PRL-2016-Cao}
Y. Cao, Z. Yang, M. Kareev, X. Liu, D. Meyers, S. Middey, D. Choudhury, P. Shafer, J. Guo, J. W. Freeland, E. Arenholz, L. Gu, and J. Chakhalian, Phys. Rev. Lett. \textbf{116}, 076802 (2016).

\bibitem{APL-2018-Wen}
F. Wen, Y. Cao, X. Liu, B. Pal, S. Middey, M. Kareev, and J. Chakhalian, Appl. Phys. Lett. \textbf{112}, 122405 (2018).

\bibitem{PRL-2011-DD}
D. A. Dikin, M. Mehta, C. W. Bark, C. M. Folkman, C. B. Eom, and V. Chandrasekhar, Phys. Rev. Lett. \textbf{107}, 056802 (2011).

\bibitem{NP-2011-JB}
J. A. Bert, B. Kalisky, C. Bell, M. Kim, Y. Hikita, H. Y. Hwang, and K. A. Moler, Nat. Phys. \textbf{7}, 767 (2011).

\bibitem{NP-2011-LL}
L. Li, C. Richter, J. Mannhart, and R. C. Ashoori, Nat. Phys. \textbf{7}, 762 (2011).

\bibitem{ARMR-2014-Stemmer}
S. Stemmer and S. J. Allen, Annu. Rev. Mater. Res. \textbf{44}, 151 (2014).

\bibitem{APLM-2015-Ahn}
K. Zou, S. Ismail-Beigi, K. Kisslinger, X. Shen, D. Su, F. J. Walker, and C. H. Ahn, APL Mater. \textbf{3}, 036104 (2015).

\bibitem{APL-2015-Liu}
X. Liu, D. Choudhury, Y. Cao, M. Kareev, S. Middey, and J. Chakhalian, Appl. Phys. Lett. \textbf{107}, 191602 (2015).

\bibitem{PRB-1997-Tokura}
T. Katsufuji, Y. Taguchi, and Y. Tokura, Phys. Rev. B \textbf{56}, 10145 (1997).

\bibitem{ACSAMI-2015-TS}
T. Sarkar, K. Gopinadhan, J. Zhou, S. Saha, J. M. D. Coey, Y. P. Feng, Ariando, and T. Venkatesan, ACS Appl. Mater. Interfaces \textbf{7}, 24616 (2015).

\bibitem{JSSC-1979-Greedan}
D. A. MacLean, H. N. Ng, J. E. Greedan, J. Solid State Chem. \textbf{30}, 35-44 (1979).

\bibitem{JSSC-2003-Eng}
H. W. Eng, P. W. Barnes, B. M. Auer, and P. M. Woodward, J. Solid State Chem. \textbf{175}, 94 (2003).

\bibitem{JSSC-2005-Ali}
R. Ali and M. Yashima, J. Solid State Chem. \textbf{178}, 2867 (2005).

\bibitem{APL-2016-Cao}
Y. Cao, S. Y. Park, X. Liu, D. Choudhury, S. Middey, D. Meyers, M. Kareev, P. Shafer, E. Arenholz, and J. Chakhalian, Appl. Phys. Lett. \textbf{109}, 152905 (2016).

\bibitem{ASS-2018-Mu}
S. Muff, M. Fanciulli, A. P. Weber, N. Pilet, Z. Risti, Z. Wang, N. C. Plumb, M. Radovi, and J. H. Dil, Appl. Surf. Sci. \textbf{432}, 41 (2018).

\end{thebibliography}
\end{document}